\documentclass[preprint,showpacs,preprintnumbers,amsmath,amssymb
              ,prb,eqsecnum]{revtex4}

\usepackage{graphicx}
\usepackage{dcolumn}
\usepackage{bm}

\begin{document}

\preprint{l-edge-magnon}

\title{Magnetic excitations in L-edge resonant inelastic\\
x-ray scattering from one-dimensional cuprates}


\author{Jun-ichi Igarashi}%
\affiliation{%
Faculty of Science, Ibaraki University, Mito, Ibaraki 310-8512, Japan}

\author{Tatsuya Nagao}
\affiliation{%
Faculty of Engineering, Gunma University, Kiryu, Gunma 376-8515, Japan}

\date{\today}

\begin{abstract}

We study the magnetic excitation spectra of
$L$-edge resonant inelastic x-ray scattering (RIXS) from the spin singlet 
ground state in one-dimensional undoped cuprates.
Analyzing the transition amplitudes of the magnetic excitations
in the second-order dipole allowed process, we find that
the magnetic excitations are brought about not only on the core-hole
site but also on the neighboring sites.
The RIXS spectra are expressed by the one-spin correlation function 
in the scattering channel with changing polarization,
and the two-spin correlation function in the scattering
channel without changing polarization.
The latter could not be brought about within the so-called UCL 
approximation.
We calculate 
these correlation functions
on a finite-size ring. 
An application to the possible RIXS spectra in Sr$_2$CuO$_3$ demonstrates 
that the contribution of the two-spin correlation function could be
larger than that of the one-spin correlation function in the $\sigma$
polarization for the momentum transfer around the zone center.

\end{abstract}

\pacs{78.70.Ck, 72.10.Di, 78.20.Bh, 74.72.Cj} 
\maketitle

\section{\label{sect.1}Introduction}

Resonant inelastic x-ray scattering (RIXS) has attracted
much interest as a useful tool to investigate excited 
states in solids.
Both the $K$-edge and the $L$-edge resonances are available  
in transition-metal compounds. To observe the momentum dependence of 
the spectra, it is suitable to use the $K$-edge resonance that
the $1s$-core electron is prompted to an empty $4p$ state
by absorbing photon and then the photo-excited $4p$ electron is
recombined with the core hole by emitting photon.
It is known that the intensities for energy loss $\omega$ above
several eV's arise from charge excitations to screen the core-hole
potential.
\cite{Kao96,Hill98,Hasan00,Kim02,Inami03,Kim04-1,Suga05,
Tsutsui99,Okada06,Nomura04,Nomura05,Igarashi06,vdBrink06,Ament07}
A useful formalism has been developed
\cite{Nomura04,Nomura05,Igarashi06} 
on the basis of the Keldysh Green function,\cite{Keldysh65}
in which the RIXS spectra are described in terms of the $3d$-density-density 
correlation function. The spectra have been calculated
for undoped cuprates,\cite{Nomura04,Nomura05,Igarashi06}
NiO,\cite{Takahashi07} and LaMnO$_3$,\cite{Semba08}
in good agreement with the experiments.
In addition to such spectra, the intensities for $\omega$ below 1 eV have been 
observed from the $K$-edge RIXS in La$_2$CuO$_4$.\cite{Hill08,Ellis10}
Their origin is attributed to the magnetic excitations
caused by the modification of the exchange coupling 
under the presence of the core-hole potential.\cite{vdBrink05,vdBrink07}
The above-mentioned formalism has been adapted to describe such 
magnetic excitations,
having led to the expression that the RIXS spectra are proportional 
to the two-spin correlation function, which has been evaluated
by the $1/S$-expansion method.\cite{Nagao07}

Recently, the L-edge RIXS experiments with high resolutions have been carried 
out in transition-metal compounds.
The observed spectra for $\omega$
around several hundreds meV's can be understood as coming from magnetic
excitations.\cite{Braicovich09,Braicovich10,Guarise10}
The $L$-edge resonance in the undoped cuprates is described by
the process that the $2p$-core electron is prompted to the empty
$x^2-y^2$ orbital by absorbing photon, and then an occupied $3d$ electron
combines with the core hole by emitting photon.
If the $3d$ orbital in the photo-emitting process is different from the one
in the photo-absorbing process, the excitations within the $3d$ states
are brought about (``d-d" excitation).\cite{Ghiringhelli04}
Even if the $3d$ orbitals in the photo-absorbing and photo-emitting processes 
are the same, purely magnetic excitations could be generated 
when the direction of the staggered moment deviates from the $z$ direction.
\cite{Ament09} 
The spectra observed as a function of energy loss exhibit 
a systematic variation with changing momentum transfer.\cite{Guarise10}
That is, the peak position moves according to the spin-wave dispersion curve,
while their shapes exhibit structures indicative of two- or three-magnon 
excitations.

In our previous paper,\cite{Igarashi2011} we have analyzed the process 
generating the final states in the second-order dipole allowed process,
and have clarified how the spin-flip excitations are brought about on
the core-hole site. We have also found that the spin-conserving
excitations are brought about around the core-hole site
due to the strong perturbation working in the intermediate state.
Note that, within the conventional approach called UCL approximation,
\cite{Ament09} no spin-conserving excitation is possible, and 
the spin-flip excitation is generated just on the core-hole site.
Since no core hole exists in the final state, these excitations 
are freely moving in crystal.
We have treated the quantum fluctuation in the final state
within the $1/S$ expansion method, which is known to work well in the 
two-dimensional $S=1/2$ Heisenberg antiferromagnet.\cite{Igarashi05}
On the basis of these results, we have analyzed the RIXS spectra
in Sr$_2$CuO$_2$Cl$_2$. 
The result shows that substantial intensities are generated in
the high energy side of the one-magnon peak, which are originated 
from two- and three-magnon excitations, 
in good agreement with the experiment.\cite{Guarise10}

The above analysis has been carried out in the antiferromagnetic ordered state.
Although the spin-flip excitations are likely to extend on neighboring sites
due to the second-order process, the systematic analysis of such effects is 
quite difficult there, because the spherical symmetry is broken in the 
spin space. 
On the other hand, in the one-dimensional Heisenberg 
antiferromagnet, since the ground state is the spin singlet,
the spherical symmetry remains intact in the spin space.
Such a system is quite suitable to study how the spin excitations 
are generated around the core-hole site.
The purpose of this paper is to clarify the nature of the magnetic excitations 
in RIXS by studying the spectra on the one-dimensional system.
We analyze the second-order process on a microscopic model,
with distinguishing the spin-flip and spin-conserving excitations.
As a result, the RIXS spectra are expressed by 
the one-spin and two-spin correlation functions
in the scattering channels with and without changing polarization,
respectively.
Note that the former includes the effect 
of the magnetic excitations generated on the neighboring sites. 
We could evaluate convincingly the amplitudes leading to these correlation
functions on a finite-size ring.
The amplitudes of the excitations on the neighboring sites are found to
increase with decreasing values of
the life-time broadening width of the core hole.
The present analysis contrasts with the previous analysis
by Haverkort,\cite{Haverkort2010}
where the form of the magnetic excitations is inferred from 
Hannon's formula\cite{Hannon1988} of the resonant elastic scattering.
The analysis as well as the UCL approximation applied to the one-dimensional
system\cite{Forte2011} could not describe the extended nature of the 
excitations.  Finally we analyze possible RIXS spectra in Sr$_2$CuO$_3$,
demonstrating that the contribution of the two-spin correlation function
becomes larger than that of the one-spin correlation function in the $\sigma$
polarization, and could be distinguished by comparing the spectral shape 
as a function of energy loss between in the $\sigma$ and $\pi$ polarizations.

The present paper is organized as follows.
In Sec. \ref{sect.2}, we describe the Hamiltonian responsible for magnetic
excitations and transition-matrix elements relevant to the $L_{2,3}$ edges
in cuprate compounds.
We analyze the second-order process giving rise to the magnetic
excitations through the intermediate state, and express the RIXS spectra 
in terms of spin-correlation functions. The amplitudes leading to 
the spin-correlation functions are evaluated on a finite-size ring.
In Sec. \ref{sect.3}, we numerically calculate the spectra on a finite-size
chain. Section \ref{sect.4} is devoted to the concluding remarks.
In Appendix, the $L_{2,3}$-edge absorption coefficient is briefly
discussed.

\section{\label{sect.2}Formulation of RIXS spectra at the $L_{2,3}$-edge}

\subsection{Second-order dipole allowed process}

Aiming at the application to the one-dimensional cuprates
such as Sr$_2$CuO$_3$ and SrCuO$_2$, we consider each Cu atom has 
one hole in the $x^2-y^2$ orbital at the half-filling,
where the $x$ and $y$ axes are defined
along the Cu-O bonds and the $z$ along the crystal $c$ axis.
The low-energy spin excitations are described by
the one-dimensional antiferromagnetic Heisenberg Hamiltonian,
\begin{equation}
 H_{\rm mag}=J \sum_{\langle i,j\rangle} 
                   \textbf{S}_i\cdot \textbf{S}_{j},
\end{equation}
with $S=1/2$ and $J>0$.
The ground state is known to be the spin singlet due to the
quantum fluctuation. Unlike the antiferromagnetic ordered state,
the spin singlet state possesses no special direction in the spin space.
To manifest the rotational invariance in the spin space, 
we define the spin coordinate frame of $x'$, $y'$, $z'$ axes by rotating
the crystal-fixed coordinate frame of $a$, $b$, $c$ axes
with the Euler angles $\alpha$, $\beta$, and $\gamma$.

In the electric dipole ($E$1) transition, a $2p$-core electron is excited to 
the $3d$ states at the transition-metal $L_{2,3}$-edge.
The $2p$ states are characterized by the 
total angular momentum $j=3/2$ and $1/2$
due to the strong spin-orbit interaction. 
The eigenstates with $j=3/2$ may be expressed as
$|\phi_{1}\uparrow\rangle$, 
$\sqrt{1/3}|\phi_{1}\downarrow\rangle+\sqrt{2/3}|\phi_{0}\uparrow\rangle$, 
$\sqrt{2/3}|\phi_{0}\downarrow\rangle+\sqrt{1/3}|\phi_{-1}\uparrow\rangle$,
$|\phi_{-1}\downarrow\rangle$, for $m=3/2$, $1/2$, $-1/2$, $-3/2$, 
respectively, and those with $j=1/2$ may be expressed as
$-\sqrt{2/3}|\phi_{1}\downarrow\rangle+\sqrt{1/3}|\phi_{0}\uparrow\rangle$, 
$-\sqrt{1/3}|\phi_{0}\downarrow\rangle+\sqrt{2/3}|\phi_{-1}\uparrow\rangle$, 
for $m=1/2$ and $-1/2$, respectively,
where $m$ represents the magnetic quantum number. 
The orbitals $\phi_{1}$, $\phi_{0}$, and $\phi_{-1}$ 
have the same angular dependence as the spherical harmonics 
$Y_{11}$, $Y_{10}$ and $Y_{1-1}$, respectively.
The coordinate frame for both orbital and spin of the core hole
is fixed to the crystal, which is different from that for the spins of 
the $3d$ states, that is, $\uparrow$ and $\downarrow$ are 
associated with the direction of the crystal $c$ axis.
The corresponding interaction between photon and electron at site $i$ 
may be described as
\begin{equation}
H_{\rm int}=w\sum_{\textbf{q}, \mu}
\frac{1}{\sqrt{2\omega_{\textbf{q}}}}
\sum_{i, m, \sigma}D^{\mu}(jm,\sigma)
h_{jm}^{\dagger}c_{\textbf{q}\mu}d_{i\sigma}
{\rm e}^{i\textbf{q}\cdot\textbf{r}_{i}} +{\rm H.c.},
\label{eq.tran1}
\end{equation}
where $c_{{\bf q}\mu}$ stands for the annihilation operator of photon 
with momentum ${\bf q}$ and polarization $\mu$.
The $h_{jm}^{\dagger}$ represents the creation operator of the $2p$ hole with
$jm$, and $d_{i\sigma}$ denotes the annihilation operator of the $3d$ hole
with the $x^2-y^2$ orbital and spin $\sigma$.
The $w$ is a constant proportional to 
$\int_0^{\infty}r^3R_{3d}(r)R_{2p}(r){\rm d}r$
where $R_{3d}(r)$ and $R_{2p}(r)$ are the radial wave-functions for 
the $3d$ and $2p$ states of Cu atom. The $D^{\mu}(jm,\sigma)$ describes
the dependence on the core-hole state and $3d$ spin,
which is shown in Table I of Ref. \onlinecite{Igarashi2011}.

With this interaction, the RIXS spectra may be expressed by the second-order
dipole allowed process:
\begin{eqnarray}
 W(q_f\alpha_f;q_i\alpha_i)
 &=& 2\pi\sum_{f'}\left|\sum_{n}
  \frac{\langle \Phi_{f'}|H_{\rm int}|n\rangle
        \langle n|H_{\rm int}|\Phi_i\rangle}
       {E_g+\omega_i-E_n} \right|^2 \nonumber\\
 &\times&\delta(E_g+\omega_i-E_{f'}-\omega_f),
\label{eq.optical}
\end{eqnarray} 
with $q_i\equiv({\bf q}_i,\omega_i)$, $q_f\equiv({\bf q}_f,\omega_f)$, 
$|\Phi_i\rangle = c_{q_i\alpha_i}^{\dagger}|g\rangle$,
$|\Phi_{f'}\rangle=c_{q_f \alpha_f}|f'\rangle$,
where $|g\rangle$ and $|f'\rangle$ represent the ground state and
excited states of the matter with energy $E_g$ and $E_{f'}$, respectively.

\subsection{Magnetic excitations around the core-hole site}

Assuming that the core hole is created at the origin in the intermediate state, 
we write the ground state $|g\rangle$ of $H_{\rm mag}$ as
\begin{equation}
 |g\rangle = |\uparrow\rangle|\psi_0^{\uparrow}\rangle
           + |\downarrow\rangle|\psi_0^{\downarrow}\rangle,
\end{equation}
where $|\uparrow\rangle$ and $|\downarrow\rangle$ represent the spin
states at the origin, and $|\psi_0^{\uparrow}\rangle$ and 
$|\psi_0^{\downarrow}\rangle$ are constructed 
by the bases of the rest of spins. 

Then, just after the $E1$ transition, the wave function may be written as
\begin{equation}
 H_{\rm int}|g\rangle \propto\sum_{m}\left[ 
\sum_{\sigma=\uparrow,\downarrow}
    D^{\alpha_i}(jm,\sigma)|\psi_0^{\sigma}\rangle
  \right] |jm\rangle,
\end{equation}
where $|jm\rangle$ represents the core hole state.
Note that the spin degree of freedom is lost at the origin in the 
intermediate state.

Let $H'$ be the Hamiltonian in the intermediate state.
The states $|\psi_0^{\uparrow}\rangle$ and 
$|\psi_0^{\downarrow}\rangle$ are not the eigenstates of $H'$. 
Introducing the normalized eigenstate $|\phi_{\eta}\rangle$'s of $H'$ 
with eigenvalue $\epsilon'_{\eta}$, we rewrite the second-order optical
process as
\begin{eqnarray}
 &&\sum_n H_{\rm int}|n\rangle\frac{1}{\omega_i+E_g-E_n}
   \langle n|H_{\rm int}|g\rangle \nonumber \\
   &\propto& \sum_{m, \sigma, \sigma'} D^{\alpha_f}(jm,\sigma)^{*}
                             D^{\alpha_i}(jm,\sigma') \nonumber \\
& & \times
   \sum_{\eta}|\sigma\rangle |\phi_{\eta}\rangle R(\epsilon'_{\eta})
    \langle\phi_{\eta}|\psi_{0}^{\sigma'}\rangle,
\label{eq.process1}
\end{eqnarray}
with
\begin{equation}
 R(\epsilon'_{\eta}) =
 \frac{1}{\omega_i+\epsilon_g -\epsilon_{\textrm{core}}+i\Gamma - \epsilon'_{\eta}},
\end{equation}
where $\epsilon_g$ represents the ground state energy of $H_{\rm mag}$.
The $\epsilon_{\textrm{core}}$ denotes 
the energy required to create a core hole in the state 
$|jm\rangle$ and the $3d^{10}$-configuration.
The $\Gamma$ stands for the life-time broadening width of the core hole;
$\Gamma\sim 0.3$ eV at the Cu $L_3$ edge. We write the polarization-dependent 
factor by 

\begin{eqnarray}
\sum_{m}D^{\alpha_f}(jm,\sigma)^{*}D^{\alpha_i}(jm,\sigma)
&\equiv& P_{\sigma}^{(0)}(j;\alpha_f,\alpha_i), \\
\sum_{m}D^{\alpha_f}(jm,\sigma)^{*}D^{\alpha_i}(jm,-\sigma) 
&\equiv& P_{\sigma}^{(1)}(j;\alpha_f,\alpha_i).
\end{eqnarray}

Table \ref{table.2} shows $P_{\sigma}^{(0)}$ and 
$P_{\sigma}^{(1)}$ for $\alpha_i$ and
$\alpha_f$ along the $x$, $y$, and $z$ axes,
where $\textrm{sgn}(\sigma)=1$ for $\sigma=\uparrow$ and $-1$ for 
$\sigma=\downarrow$.

\begin{table*}
\caption{\label{table.2}
$P_{\sigma}^{(0)}(j;\alpha_f,\alpha_i)$ and 
$P_{\sigma}^{(1)}(j;\alpha_f,\alpha_i)$ 
where upper and lower signs correspond to $\sigma=\uparrow$ and $\downarrow$, 
respectively.
}
\begin{ruledtabular}
\begin{tabular}{rcrrrrrrr}
      & & $P_{\sigma}^{(0)}$ & & & & $P_{\sigma}^{(1)}$ & & \\
\hline
$j$   & $\alpha_f \setminus \alpha_i$ & $x$  & $y$ & $z$ &  & $x$ & $y$ & $z$ \\ 
\hline
$\frac{3}{2}$ & $x$ & $\frac{2}{15}$ & $\mp \frac{i}{15}\cos\beta$ & $0$ & 
& $0$ & $\frac{i}{15} \textrm{e}^{\pm i \gamma} \sin\beta$ & $0$ \\
      & $y$ & $\pm \frac{i}{15}\cos\beta$ & $\frac{2}{15}$ & $0$ & & $-\frac{i}{15} \textrm{e}^{\pm i \gamma} \sin\beta$ & $0$ & $0$ \\
      & $z$ & $0$ & $0$ & $0$ & & $0$ & $0$ & $0$ \\
\hline
$\frac{1}{2}$ & $x$ & $\frac{1}{15}$ & $\pm \frac{i}{15}\cos\beta$ 
& $0$ & & $0$ & 
$-\frac{i}{15} \textrm{e}^{\pm i \gamma} \sin\beta$ & $0$ \\ 
      & $y$ & $\mp \frac{i}{15}\cos\beta$
 & $\frac{1}{15}$ & $0$ & & 
$\frac{i}{15} \textrm{e}^{\pm i \gamma} \sin\beta$ & $0$ & $0$ \\
      & $z$ & $0$ & $0$ & $0$ & & $0$ & $0$ & $0$ \\
\end{tabular}
\end{ruledtabular}
\end{table*}

\subsubsection{Scattering channel with changing polarization}

We analyze the scattering channel that changes the polarization
during the process at the $L_3$-edge.
Let $\alpha_i$ and $\alpha_f$ be along $y$ and $x$ axes, respectively.  
We have the spin-conserving term coming from $P_{\sigma}^{(0)}$ and
the spin-flipping term coming from $P_{\sigma}^{(1)}$.
The spin-conserving term is given by
\begin{widetext}
\begin{eqnarray}
 &&\sum_n H_{\rm int}|n\rangle\frac{1}{\omega_i+E_g-E_n}
   \langle n|H_{\rm int}|g\rangle \nonumber \\
 &\propto& P_{\uparrow}^{(0)}
  |\uparrow\rangle\sum_{\eta}|\phi_{\eta}\rangle R(\epsilon'_{\eta})
  \langle\phi_{\eta}|\psi_{0}^{\uparrow}\rangle 
 + P_{\downarrow}^{(0)}
  |\downarrow\rangle\sum_{\eta}|\phi_{\eta}\rangle R(\epsilon'_{\eta})
  \langle\phi_{\eta}|\psi_{0}^{\downarrow}\rangle, \nonumber\\
 &\propto& \left(-\frac{i}{15}\right)\cos\beta \left\{
  |\uparrow\rangle\sum_{\eta}|\phi_{\eta}\rangle R(\epsilon'_{\eta})
  \langle\phi_{\eta}|\psi_{0}^{\uparrow}\rangle 
  - |\downarrow\rangle\sum_{\eta}|\phi_{\eta}\rangle R(\epsilon'_{\eta})
  \langle\phi_{\eta}|\psi_{0}^{\downarrow}\rangle \right\}. 
\label{eq.non-flip}
\end{eqnarray}
\end{widetext}
We consider the states $S_{0}^{z'}|g\rangle$ and 
$(S_{1}^{z'}+S_{-1}^{z'})|g\rangle$
onto which Eq.~(\ref{eq.non-flip}) is projected.
Although they are orthogonal to $|g\rangle$, they are not orthogonal 
to each other nor normalized.
Let $|\psi_1\rangle$ and $|\psi_2\rangle$ be $S_{0}^{z'}|g\rangle$ and
$(S_{1}^{z'}+S_{-1}^{z'})|g\rangle$, respectively. Then, the overlap matrix
$(\hat{\rho}_{z'})_{i,j}\equiv\langle\psi_i|\psi_j\rangle$ is given by
\begin{equation}
 \hat{\rho}_{z'}= \left(\begin{array}{cc}
       \frac{1}{4} & 2a \\
       2a & \frac{1}{2}+2b
       \end{array} \right), 
\end{equation}
where $a$ and $b$ are static correlation functions between the nearest and
the second-nearest pairs, respectively, that is,
$a=\langle S_{1}^{z'}S_{0}^{z'}\rangle=\langle S_{-1}^{z'}S_{0}^{z'}\rangle$,
$b=\langle S_{1}^{z'}S_{-1}^{z'}\rangle$.
Using the inverse of $\hat{\rho}_{z'}$, we obtain
\begin{eqnarray}
 &&\sum_n H_{\rm int}|n\rangle\frac{1}{\omega_i+E_g-E_n}
   \langle n|H_{\rm int}|g\rangle \nonumber\\
 &\propto& \left(-\frac{i}{15}\right)
 \cos\beta \left\{
   f_{1}^{(1)}(\omega_i) S_{0}^{z'}
  +f_{2}^{(1)}(\omega_i) (S_{1}^{z'}+S_{-1}^{z'})
 \right\}|g\rangle,
\end{eqnarray}
where 
\begin{equation}
  f_{m}^{(1)}(\omega_i)=\sum_{n=1,2}(\hat{\rho}_{z'}^{-1})_{m,n}(Q_{z'}^{(1)})_{n},
\end{equation}
with
\begin{eqnarray}
(Q_{z'}^{(1)})_{1} &=& \langle\psi_{0}^{\uparrow}|
\sum_{\eta}|\phi_{\eta}\rangle R(\epsilon'_{\eta})\langle\phi_{\eta}|
        \psi_{0}^{\uparrow}\rangle ,\\
(Q_{z'}^{(1)})_{2} &=& 4\langle\psi_{0}^{\uparrow}|S_{1}^{z'}
\sum_{\eta}|\phi_{\eta}\rangle R(\epsilon'_{\eta})\langle\phi_{\eta}|
        \psi_{0}^{\uparrow}\rangle .
\end{eqnarray}
Therefore, the final state is expressed as 
\begin{eqnarray}
 &&\left(-\frac{i}{15}\right)\cos\beta 
\left[f_{1}^{(1)}(\omega_i)S_0^{z'}
     +f_{2}^{(1)}(\omega_i)(S_{1}^{z'}+S_{-1}^{z'})\right]
   |g\rangle \nonumber\\
&=&\left(-\frac{i}{15}\right)
 \mbox{\boldmath{$\alpha$}}_{f\perp}\times
 \mbox{\boldmath{$\alpha$}}_{i\perp}\cdot
 \left[f_{1}^{(1)}(\omega_i){\bf S}_{0\parallel}
     +f_{2}^{(1)}(\omega_i)({\bf S}_{1\parallel}+{\bf S}_{-1\parallel})\right]
   |g\rangle 
\label{eq.sparallel}
\end{eqnarray}
where ${\bf S}_{j \parallel}$ stands for the $z'$ component of 
${\bf S}_{j}$. 
The $\mbox{\boldmath{$\alpha$}}_{i\perp}$ and
$\mbox{\boldmath{$\alpha$}}_{f\perp}$ represent the polarization vectors
projected onto the $x$-$y$ plane. Therefore 
$\mbox{\boldmath{$\alpha$}}_{f\perp}\times
 \mbox{\boldmath{$\alpha$}}_{i\perp}$
is always parallel to the $z$ axis.

The spin-flip term may be expressed as
\begin{eqnarray}
  &&\sum_n H_{\rm int}|n\rangle\frac{1}{\omega_i+E_g-E_n}
   \langle n|H_{\rm int}|g\rangle \nonumber \\
  &\propto& P_{\downarrow}^{(1)}
  |\downarrow\rangle\sum_{\eta}|\phi_{\eta}\rangle R(\epsilon'_{\eta})
  \langle\phi_{\eta}|\psi_{0}^{\uparrow}\rangle
  + P_{\uparrow}^{(1)}
  |\uparrow\rangle\sum_{\eta}|\phi_{\eta}\rangle R(\epsilon'_{\eta})
  \langle\phi_{\eta}|\psi_{0}^{\downarrow}\rangle, \nonumber\\
  &\propto& \left(\frac{i}{15}\right)
  \sin\beta \left\{{\rm e}^{-i\gamma}
   |\downarrow\rangle\sum_{\eta}|\phi_{\eta}\rangle R(\epsilon'_{\eta})
   \langle\phi_{\eta}|\psi_{0}^{\uparrow}\rangle
   + {\rm e}^{i\gamma}|\uparrow\rangle\sum_{\eta}|\phi_{\eta}\rangle
  R(\epsilon'_{\eta})\langle\phi_{\eta}|\psi_{0}^{\downarrow}\rangle \right\}.
\label{eq.spin-flip}
\end{eqnarray}
We project this state onto a pair of states $S_0^{+}|g\rangle$ and 
$(S_{1}^{+}+S_{-1}^{+})|g\rangle$, and another pair of states 
$S_0^{-}|g\rangle$ and $(S_{1}^{-}+S_{-1}^{-})|g\rangle$, where
$S_{j}^{\pm}=S_{j}^{x'}\pm i S_{j}^{y'}$. 
The states belonging to different pairs are orthogonal.
Defining the overlap matrices $\hat{\rho}_{+}$ and $\hat{\rho}_{-}$
for each pair in a similar way used for $\hat{\rho}_{z '}$, 
we immediately notice that
\begin{equation}
 \hat{\rho}_{+}=\hat{\rho}_{-}=2\hat{\rho}_{z '} .
\end{equation}
The second term of Eq.~(\ref{eq.spin-flip}) is projected onto 
the first pair of the states, leading to
\begin{equation}
 \propto \left(\frac{i}{15}\right)
 \sin\beta {\rm e}^{i\gamma}
\left\{g_{1}(\omega_i)S_0^{+}
+g_{2}(\omega_i)(S_{1}^{+}+S_{-1}^{+})\right\}|g\rangle,
\end{equation}
where
\begin{equation}
  g_{m}(\omega_i)=\sum_{n=1,2}(\hat{\rho}_{+}^{-1})_{m,n}(Q_{+}^{(1)})_{n},
\end{equation}
with
\begin{eqnarray}
(Q_{+}^{(1)})_{1} &=& \langle\psi_{0}^{\downarrow}|
\sum_{\eta}|\phi_{\eta}\rangle R(\epsilon'_{\eta})\langle\phi_{\eta}|
        \psi_{0}^{\downarrow}\rangle = (Q_{z}^{(1)})_{1} ,\\
(Q_{+}^{(1)})_{2} &=& 2\langle\psi_{0}^{\uparrow}|S_{1}^{-}
\sum_{\eta}|\phi_{\eta}\rangle R(\epsilon'_{\eta})\langle\phi_{\eta}|
        \psi_{0}^{\downarrow}\rangle = (Q_{z}^{(1)})_{2} . 
\end{eqnarray}
Hence we have $g_{1}(\omega_i)=(1/2)f_{1}^{(1)}(\omega_i)$ and
$g_{2}(\omega_i)=(1/2)f_{2}^{(1)}(\omega_i)$.
The similar relations are obtained for the first term of 
Eq.~(\ref{eq.spin-flip}) by using the second pair of the states. 
Therefore Eq.~(\ref{eq.spin-flip}) is expressed as
\begin{eqnarray}
  &\propto&\left(\frac{i}{15}\right) \sin\beta 
   \Bigl\{\cos\gamma [f_{1}^{(1)}(\omega_i)S_{0}^{x'}
           +f_{2}^{(1)}(\omega_i)(S_{1}^{x'}+S_{-1}^{x'})]\nonumber\\
       &-&\sin\gamma [f_{1}^{(1)}(\omega_i)S_{0}^{y'}
           +f_{2}^{(1)}(\omega_i)(S_{1}^{y'}+S_{-1}^{y'})]
   \Bigr\}|g\rangle \nonumber\\
  &=& \left(-\frac{i}{15}\right) 
    \mbox{\boldmath{$\alpha$}}_{f\perp}\times
    \mbox{\boldmath{$\alpha$}}_{i\perp}\cdot
    [f_{1}^{(1)}(\omega_i){\bf S}_{0\perp}
    +f_{2}^{(1)}(\omega_i)({\bf S}_{1\perp}+{\bf S}_{-1\perp})] 
   |g\rangle ,
\end{eqnarray}
where ${\bf S}_{j \perp}$ stands for the component perpendicular to
the $z'$ axis of ${\bf S}_{j}$.
Combining this result with Eq,~(\ref{eq.sparallel}), we
finally obtain
\begin{eqnarray}
  &&\sum_n H_{\rm int}|n\rangle\frac{1}{\omega_i+E_g-E_n}
   \langle n|H_{\rm int}|g\rangle \nonumber\\
   &=& \left(-\frac{i}{15}\right)
   \mbox{\boldmath{$\alpha$}}_f\times
   \mbox{\boldmath{$\alpha$}}_i\cdot 
    [f_{1}^{(1)}(\omega_i){\bf S}_{0}
    +f_{2}^{(1)}(\omega_i)({\bf S}_{1}+{\bf S}_{-1})]|g\rangle.
\end{eqnarray}
Excitations on further neighboring sites are included by adding a term 
$f_{3}^{(1)}(\omega_i)({\bf S}_{2}+{\bf S}_{-2})|g\rangle$
with $f_{3}^{(1)}(\omega_i)$ similarly determined coefficient.

Collecting up the amplitudes from all sites,
we obtain the expression of the RIXS spectra for the  polarizations 
$\mbox{\boldmath{$\alpha$}}_{i(f)}
=(\alpha_{i(f)}^{x},\alpha_{i(f)}^{y},\alpha_{i(f)}^{z})$,
\begin{equation}
 W(q_f,\omega_f,\alpha_f;q_i,\omega_i,\alpha_i) = 
 \frac{w^4}{4\omega_i\omega_f}\left(\frac{1}{15}\right)^2
 \left(\alpha_f^{x}\alpha_i^{y}-\alpha_f^{y}\alpha_i^{x}\right)^2
                  Y^{(1)}(\omega_i;q,\omega),
\label{eq.rixs.flip}
\end{equation}
where
\begin{equation}
 Y^{(1)}(\omega_i;q,\omega)=\int\langle Z^{(1)\dagger}(\omega_i;q,t) 
  Z^{(1)}(\omega_i;q,0)\rangle {\rm e}^{i\omega t}{\rm d}t,
\label{eq.y1}
\end{equation}
with
\begin{eqnarray}
 Z^{(1)}(\omega_i;q)&=&\sum_{j}
  [f_{1}^{(1)}(\omega_i) S_{j}^{z}
    +f_{2}^{(1)}(\omega_i)(S_{j+1}^{z}+S_{j-1}^{z})
    +f_{3}^{(1)}(\omega_i)(S_{j+2}^{z}+S_{j-2}^{z})]
   {\rm e}^{-iqr_j} \nonumber\\
 &=& \bigl[f_{1}^{(1)}(\omega_i)+2f_{2}^{(1)}(\omega_i)\cos q
     +2f_{3}^{(1)}(\omega_i)\cos (2q)\bigr]
     \sum_{j}S_{j}^{z} {\rm e}^{-iqr_j}.
\end{eqnarray}
The presence of $f_{2}^{(1)}(\omega_i)$ and $f_{3}^{(1)}(\omega_i)$ 
modifies the $q$-dependence
of the spectra, but the spectral shape as a function of $\omega$ is
expressed by the conventional correlation function.

\subsubsection{Scattering channel without changing polarization}

In this scattering channel, only the spin-conserving excitations are brought 
about through the diagonal components of $P_{\sigma}^{(0)}$. For each component,
we have
\begin{eqnarray}
 &&\sum_n H_{\rm int}|n\rangle\frac{1}{\omega_i+E_g-E_n}
   \langle n|H_{\rm int}|g\rangle \nonumber \\
 &\propto& \left(\frac{2}{15}\right) \left\{
  |\uparrow\rangle\sum_{\eta}|\phi_{\eta}\rangle R(\epsilon'_{\eta})
  \langle\phi_{\eta}|\psi_{0}^{\uparrow}\rangle 
  + |\downarrow\rangle\sum_{\eta}|\phi_{\eta}\rangle R(\epsilon'_{\eta})
  \langle\phi_{\eta}|\psi_{0}^{\downarrow}\rangle \right\}. 
\label{eq.diago}
\end{eqnarray}

Since the polarization dependence behaves as 
$\mbox{\boldmath{$\alpha$}}_{f\perp}\cdot
\mbox{\boldmath{$\alpha$}}_{i\perp}$,
the corresponding spin excitations may be expressed by operating even number of
spin operators on the ground state. Therefore it is reasonable to assume
the excited states as 
$({\bf S}_{1}+{\bf S}_{-1})\cdot{\bf S}_{0}|g\rangle$,
$({\bf S}_{2}+{\bf S}_{-2})\cdot{\bf S}_{0}|g\rangle$, and
${\bf S}_{1}\cdot{\bf S}_{-1}|g\rangle$.
Since these states are not orthogonal to $|g\rangle$,
we define the overlap matrix $\hat{\rho}$ by including $|g\rangle$
to be projected in addition to the above three states.
The procedure of projection is the same as before.
Here we simply write down the result;
\begin{eqnarray}
  &&\sum_n H_{\rm int}|n\rangle\frac{1}{\omega_i+E_g-E_n}
   \langle n|H_{\rm int}|g\rangle \nonumber \\
   &\propto& \frac{2}{15} \mbox{\boldmath{$\alpha$}}_{f\perp}\cdot
   \mbox{\boldmath{$\alpha$}}_{i\perp}
   \Bigl\{f_{1}^{(2)}(\omega_i)+
     f_{2}^{(2)}(\omega_i)({\bf S}_{1}+{\bf S}_{-1})\cdot{\bf S}_{0}\nonumber\\
  &+&f_{3}^{(2)}(\omega_i)({\bf S}_{2}+{\bf S}_{-2})\cdot {\bf S}_{0} 
   +f_{4}^{(2)}(\omega_i){\bf S}_{1}\cdot{\bf S}_{-1}
     \Bigr\}|g\rangle.
\label{eq.two}
\end{eqnarray}

Collecting up the contribution from all sites,
we obtain the RIXS spectra for the  polarization vectors 
$\mbox{\boldmath{$\alpha$}}_i$ and
$\mbox{\boldmath{$\alpha$}}_f$, 
\begin{equation}
 W(q_f,\omega_f,\alpha_f;q_i,\omega_i,\alpha_i) = 
 \frac{w^4}{4\omega_i\omega_f}\left(\frac{2}{15}\right)^2
 \left(\alpha_{f}^{x}\alpha_{i}^{x}+\alpha_{f}^{y}\alpha_{i}^{y}\right)^2
                  Y^{(2)}(\omega_i;q,\omega),
\label{eq.rixs.nonflip1}
\end{equation}
where
\begin{equation}
 Y^{(2)}(\omega_i;q,\omega) = \int
 \langle Z^{(2)\dagger}(\omega_i;q,t)Z^{(2)}(\omega_i;q,0)\rangle 
  {\rm e}^{i\omega t}{\rm d}t,
\label{eq.y2}
\end{equation} 
with
\begin{eqnarray}
 Z^{(2)}(\omega_i;q)&=&\sum_{j}
   \Bigl\{f_{2}^{(2)}(\omega_i)({\bf S}_{j+1}+{\bf S}_{j-1})\cdot{\bf S}_{j}
    +f_{3}^{(2)}(\omega_i)({\bf S}_{j+2}+{\bf S}_{j-2})\cdot {\bf S}_{j} 
    \nonumber\\
   &+&f_{4}^{(2)}(\omega_i){\bf S}_{j+1}\cdot{\bf S}_{j-1} 
     \Bigr\}{\rm e}^{-iqr_j}.
\end{eqnarray}

In the far-off-resonance condition that 
$|\omega_i+\epsilon_g-\epsilon_{\textrm{core}}| 
\gg \Gamma, \epsilon'_{\eta}$,
and in the UCL approximation that 
$\Gamma \gg |\omega_i+\epsilon_g-\epsilon_{\textrm{core}}|,
\epsilon'_{\eta}$, we could factor out $R(\epsilon'_{\eta})$ 
in Eq.~(\ref{eq.diago}).
In such circumstances, using the relation 
$\sum_{\eta}|\phi_{\eta}\rangle\langle\phi_{\eta}|=1$, we notice that
Eq.~(\ref{eq.diago}) is proportional to $|g\rangle$, that is, no excitations
are generated.

\subsection{Evaluation of coefficients}

\begin{figure}
\includegraphics[width=8.0cm]{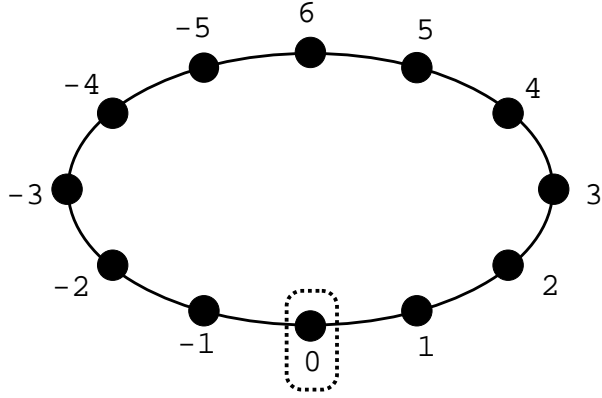}
\caption{\label{fig.ring}
A ring of 12 spins used to evaluate $f_{\mu}^{(1)}(\omega_i)$'s and
$f_{\nu}^{(2)}(\omega_i)$'s.
The spin at site 0 is annihilated in the intermediate state.
}
\end{figure}

For evaluating 
$f_{\mu}^{(1)}(\omega_i)$'s and $f_{\nu}^{(2)}(\omega_i)$'s,
we consider a system consisting of 12 spins of $S=1/2$ 
with periodic boundary conditions for the initial and final states, 
as shown in Fig.~\ref{fig.ring}. 
Since the relevant magnetic excitations are restricted around the core-hole 
site, we expect that a system having rather small size works well.
Representing $H_{\rm mag}$ by a matrix
of $924\times 924$ dimensions in the subspace of the $z '$ component of
the total spin $S_{\rm tot}^{z '}=0$, we diagonalize the Hamiltonian matrix.
We obtain the ground state energy as  $\epsilon_g/(NJ)=-0.448$, which should 
be compared with the exact value $-0.443$.\cite{Hulthen38}
On the other hand, the intermediate state is expressed by using 11 spins,
since the spin degree of freedom is lost 
at the core-hole site. Therefore $H'$ may be represented by a matrix with 
$462\times 462$ dimensions in the subspace of $S_{\rm tot}^{z '}=\pm 1/2$.

As briefly discussed in Appendix and shown there in Fig.~\ref{fig.abs},
the absorption coefficient for the transition
$2p\to 3d_{x^2-y^2}$ in cuprates has a single-peak structure;
the peak is located at 
$\omega_i = \omega_i^{0}\equiv \epsilon_{\textrm{core}}+0.8J$.
Note that the incident photon energy $\omega_i$ is usually
tuned to give the maximum absorption coefficient in RIXS experiments. 
For Sr$_2$CuO$_3$ and SrCuO$_2$, $J$ is rather large ($200-250$ meV), and 
$\Gamma\sim 0.3$ eV, and thereby we have $\Gamma/J\sim 1.2$.

Using the eigenvalues and eigenfunctions on a ring of 12 spins, we calculate
the coefficients for $\omega_i=\omega_i^{0}$.
Table \ref{table.3} lists the calculated values.
For $f_{\mu}^{(1)}$'s, $|f_{2}^{(1)}|$ and $|f_{3}^{(1)}|$ are rather
smaller than $|f_{1}^{(1)}|$ with $\Gamma/J=1.2$,
while they become larger with $\Gamma/J=0.5$, 
indicating that the effect of
magnetic excitations on neighboring sites increases with decreasing value
of $\Gamma$.
As regards $f_{\nu}^{(2)}$'s, $|f_{2}^{(2)}|$ overwhelms other absolute values
even when $\Gamma/J=0.5$. This suggests that the disturbance is nearly
limited within the nearest neighbor sites.

\begin{table*}
\caption{\label{table.3}
Coefficients $f_{\mu}^{(1)}(\omega_i^0)$'s and $f_{\nu}^{(1)}(\omega_i^0)$'s 
in units of $1/J$.
}
\begin{ruledtabular}
\begin{tabular}{cccc}
  $\Gamma/J$ & $f_{1}^{(1)}(\omega_i^0)$ & 
               $f_{2}^{(1)}(\omega_i^0)$ & 
               $f_{3}^{(1)}(\omega_i^0)$ \\ 
\hline
       $1.2$ & $(0.182,-1.752)$ & $(0.282,-0.188)$ & $(0.165,-0.106)$ \\
       $0.5$ & $(0.649,-4.395)$ & $(0.723,-0.979)$ & $(0.426,-0.567)$ \\
\hline
             &                  &                  &                  \\
\hline
  $\Gamma/J$ & $f_{2}^{(2)}(\omega_i^0)$ & 
               $f_{3}^{(2)}(\omega_i^0)$ & 
               $f_{4}^{(2)}(\omega_i^0)$ \\ 
\hline
       $1.2$ & $(0.038,-0.365)$ & $(-0.007,0.009)$ & $(0.039,-0.067)$ \\
       $0.5$ & $(-0.431,-0.543)$ & $(0.010,0.022)$ & $(-0.074,-0.142)$ \\
\end{tabular}
\end{ruledtabular}
\end{table*}

\section{\label{sect.3}RIXS spectra}

\begin{figure}
\includegraphics[width=8.0cm]{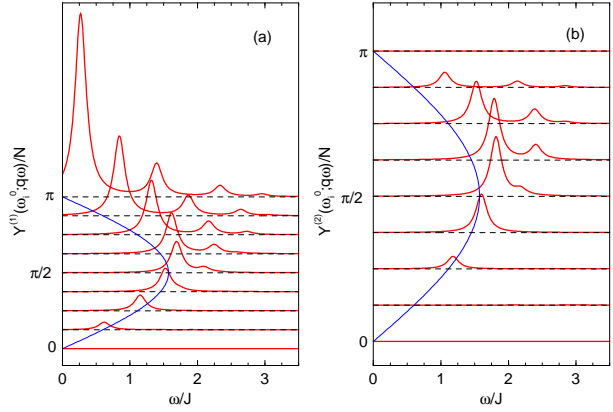}%
\caption{\label{fig.spectrum}
(Color online) Correlation functions calculated on a ring of 16 spins,
as a function of energy loss $\omega$ for various $q$-values;
(a)$Y^{(1)}(\omega_i^0;q,\omega)/N$ and (b)$Y^{(2)}(\omega_i^0;q,\omega)/N$,
with $\omega_i^{0}$ the photon energy giving rise to the peak 
in the absorption spectra, and $\Gamma/J=1.2$.
The blue lines represent the des Cloizeaux-Pearson curve,
which is the lowest boundary of the excitation energy.
}
\end{figure}

Since the magnetic excitations could propagate in the crystal in the final
state because of the absence of core hole, a small-size ring would not 
work well.
In the following, we calculate the
correlation functions on a ring of 16 spins 
from Eqs. (\ref{eq.y1}) and (\ref{eq.y2}).
Figure \ref{fig.spectrum} shows $Y^{(1)}(\omega_{i}^{0};q,\omega)$ 
and $Y^{(2)}(\omega_{i}^{0};q,\omega)$ numerically calculated 
with $\Gamma/J=1.2$.
The blue lines represent the des Cloizeaux-Pearson curve,\cite{desCloizeaux62}
$\omega=J(\pi/2)\sin q$, which is the lowest boundary of 
the excitation energy.
These functions have already been obtained from the calculation
on a finite-size system\cite{Forte2011} and also from the Bethe Ansatz 
solution.\cite{Klauser2011}
The spectral shape of $Y^{(2)}(\omega_{i}^{0};q,\omega)$ as a function of
$\omega$ seems to have more weights at higher $\omega$ than that of
$Y^{(1)}(\omega_{i}^{0};q,\omega)$.

\begin{figure}
\includegraphics[width=8.0cm]{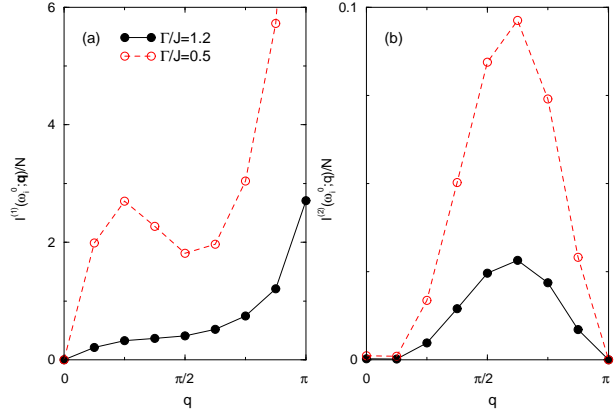}
\caption{\label{fig.intensity}
(Color online) 
Frequency-integrated intensities of the correlation functions, 
(a) $I^{(1)}(\omega_i^{0};q)/N$ and (b) $I^{(2)}(\omega_i^{0};q)/N$, 
calculated on a ring of 16 spins as a function of $q$.
The $\omega_i^{0}$ is the photon energy giving rise to the peak 
in the absorption spectra. Filled black and open red circles
correspond to $\Gamma/J=1.2$ and $0.5$, respectively.
Curves are guides to the eye.
}
\end{figure}

Figure \ref{fig.intensity} shows the integrated intensities defined by
\begin{eqnarray}
 I^{(1)}(\omega_i^{0};q) &=& \int Y^{(1)}(\omega_i^{0};q,\omega)
 \frac{{\rm d}\omega}{2\pi}, \\
 I^{(2)}(\omega_i^{0};q) &=& \int Y^{(2)}(\omega_i^{0};q,\omega)
 \frac{{\rm d}\omega}{2\pi}.
\end{eqnarray}
The $I^{(1)}(\omega_{i}^{0};q)$ vanishes with $q\to 0$, increases with 
increasing values of $q$, and remains finite with $q\to\pi$.
The presence of $f_{2}^{(1)}(\omega_i)$ and $f_{3}^{(1)}(\omega_i)$ 
makes the $q$-dependence deviate from that of the dynamical structure factor.
The deviation becomes conspicuous with $\Gamma/J=0.5$, 
because of the increase of 
$|f_{2}^{(1)}(\omega_i^0)|$ and $|f_{3}^{(1)}(\omega_i^0)|$.
The $I^{(2)}(\omega_i^{0};q)$ is found one order of magnitude smaller than
$I^{(1)}(\omega_i^{0};q)$ around the zone center.

\begin{figure}
\includegraphics[width=7.0cm]{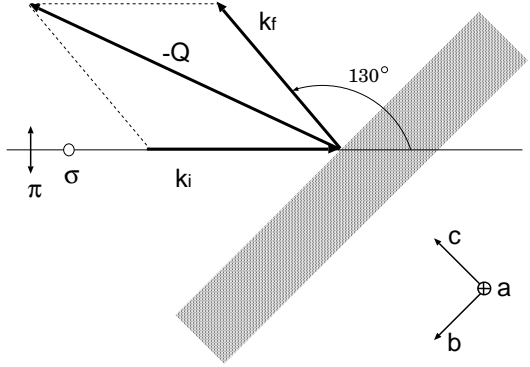}
\caption{\label{fig.geom}
Schematic view of the scattering geometry.
The scattering plane contains the $b$ and $c$ axes.
The angle between the incident and scattered x-rays is fixed 130 degrees.
}
\end{figure}

Bearing in mind a possible application to Sr$_2$CuO$_3$, 
we demonstrate the importance of $Y^{(2)}(\omega_i^0;q,\omega)$
by calculating the RIXS spectra on the $L_3$ edge
under a typical scattering geometry shown in Fig.~\ref{fig.geom},
the same geometry as used in the experiment in Sr$_2$CuO$_2$Cl$_2$; 
\cite{Guarise10}
the angle between the incident and the scattered x-ray is kept $130$ degrees,
and the scattering plane includes the $b(x)$ and $c(z)$ axes.
The polarization vector of the incident photon is then expressed as 
$\mbox{\boldmath{$\alpha$}}_i=(0,-1,0)$ for the $\sigma$ polarization and
$\mbox{\boldmath{$\alpha$}}_i=(\chi_i^{\pi},0,\tilde{\chi}_i^{\pi})$ 
for the $\pi$ polarization.
Similarly, the polarization of the scattered photon is expressed as
$\mbox{\boldmath{$\alpha$}}_f=(0,-1,0)$ for the $\sigma'$ polarization 
and
$\mbox{\boldmath{$\alpha$}}_f=(\chi_f^{\pi},0,\tilde{\chi}_f^{\pi})$ 
for the $\pi'$ polarization. 
The polarization is usually separated with the incident photon, 
but not separated with the scattered photon in experiments. 
In such a situation, we may express the RIXS spectra depending on 
the polarization of the incident photon as
\begin{equation}
I(\omega_i;q,\omega)=\frac{w^4}{4\omega_i\omega_f} 
\times 
\left\{ \begin{array}{ll}
 \left[\left(\frac{\chi_f^{\pi}}{15}\right)^2
Y^{(1)}(\omega_i;q,\omega)
     + \left(\frac{2}{15}\right)^2 Y^{(2)}(\omega_i;q,\omega)\right],
& (\sigma-{\rm pol.}), \\
 \left[\left(\frac{\chi_i^{\pi}}{15}\right)^2
Y^{(1)}(\omega_i;q,\omega)
     +\left(\frac{2\chi_f^{\pi}\chi_i^{\pi}}{15}\right)^2
  Y^{(2)}(\omega_i;q,\omega)\right],
&  (\pi-{\rm pol.}), \\
\end{array} \right. ,
\end{equation}
where $q$ is regarded as the transferred momentum projected onto the $b$ axis.
The contribution of $Y^{(2)}(\omega_i;q,\omega)$ 
relative to that of $Y^{(1)}(\omega_i;q,\omega)$ is enhanced by
$( 2/\chi_{f}^{\pi})^2$ in the $\sigma$ polarization. The contribution of
$Y^{(2)}(\omega_i;q,\omega)$ in the $\pi$
polarization is reduced from that in the $\sigma$ polarization by a factor
$(\chi_{f}^{\pi}\chi_{i}^{\pi})^2$.

\begin{figure}
\includegraphics[width=8.0cm]{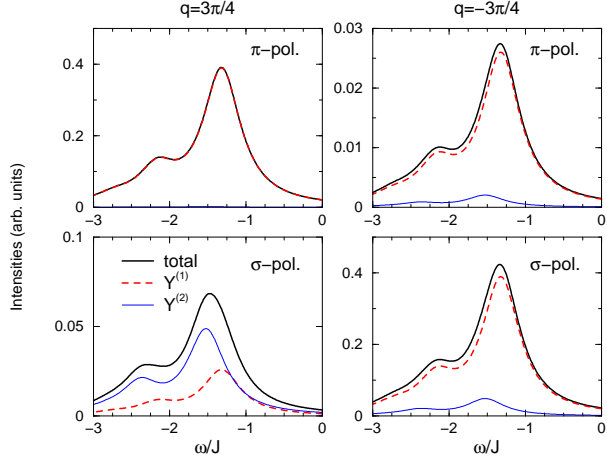}
\caption{\label{fig.RIXS}
(Color online) RIXS spectra as a function of energy loss 
$\omega$ for the momentum transfer
projected on the $b$ axis $q=3\pi/4$ and $-3\pi/4$.
Black thick solid, red thick broken, and blue thin solid lines are
total spectrum, $Y^{(1)}(\omega_i;q,\omega)$, 
and $Y^{(2)}(\omega_i;q,\omega)$, respectively.
The $\omega_i$ is set to give rise to the peak in the absorption spectra.
$J=260$ meV and $\Gamma/J=1.2$.
The calculated spectra are convoluted with the Lorentzian function with the
half width of half maximum $78$ meV. 
}
\end{figure}

Figure \ref{fig.RIXS} shows the RIXS spectrum as a function of energy loss 
$\omega$ for $q=3\pi/4(-3\pi/4)$, where $\chi_i^{\pi}=0.89(0.23)$, 
$\chi_f^{\pi}=0.23(0.89)$ for $\omega_i\sim 930$ eV and $b=3.49$ $\textrm{\AA}$.
We put $J\sim 260$ meV and $\Gamma/J=1.2$.
The calculated curves are convoluted by the Lorentzian function with
the half-width-half-maximum of the possible resolution, $78$ meV.
The contribution of $Y^{(1)}(\omega_{i}^{0};q,\omega)$ dominates the spectra 
in the $\pi$ polarization at both $q=\pm 3\pi/4$, while the contribution of 
$Y^{(2)}(\omega_i;q,\omega)$ relative to that of $Y^{(1)}(\omega_i;q,\omega)$ 
increases in the $\sigma$ polarization due to the polarization factor.

At $q=3\pi/4$, the contribution of $Y^{(2)}(\omega_{i}^{0};q,\omega)$ 
becomes even larger than that of $Y^{(1)}(\omega_{i}^{0};q,\omega)$.
Since the contribution of $Y^{(1)}(\omega_{i}^{0};q,\omega)$ in the $\sigma$ 
polarization could be estimated experimentally from the spectra 
in the $\pi$ polarization by multiplying the polarization factor, 
we may confirm from experiment the large contribution of 
$Y^{(2)}(\omega_{i}^{0};q,\omega)$ in the $\sigma$ polarization at $q=3\pi/4$. 
We also expect to observe the difference in the spectral shapes between 
in the $\pi$ and $\sigma$ polarizations,
since the spectral shape of $Y^{(2)}(\omega_{i}^{0};q,\omega)$ 
has more weights at higher $\omega$ than that of 
$Y^{(1)}(\omega_{i}^{0};q,\omega)$. 
Within the present finite-size
calculation, the spectral peak shifts to a higher energy position 
in the $\sigma$ polarization from the position in the $\pi$ polarization.
Although the system of 16 spins may be too small to discuss spectral shapes
in detail, the fact that more weights exist at higher $\omega$ in
$Y^{(2)}(\omega_i;q,\omega)$ than in $Y^{(1)}(\omega_i;q,\omega)$
has been known from the more precise calculation based on the Bethe Ansatz 
solution,\cite{Klauser2011} if $f_{3}^{(2)}(\omega_i)$ and 
$f_{4}^{(2)}(\omega_i)$ are neglected (actually they are negligible).

\section{\label{sect.4} Concluding Remarks}

We have studied the magnetic excitations in the $L$-edge RIXS 
in one-dimensional undoped cuprates.
We have analyzed the second-order dipole allowed process 
through the intermediate state, 
in which there is no spin degree of freedom at the core-hole site.
This nature of the intermediate state is found to affect strongly 
the transition amplitudes of spin excitations not only at the core-hole site
but also at neighboring sites in the final state.
This tendency is found to increase with decreasing values of $\Gamma$.
The spherical symmetry in the spin space in the ground state makes 
our analysis transparent, making it possible to analyze the amplitudes 
giving rise to excitations not only on the core-hole site but also
on the neighboring sites. We have evaluated such amplitudes in a finite-size
ring. Note that the analysis of the RIXS from the antiferromagnetic 
ordered state was complicated due to the presence of
the asymmetry in the spin space.\cite{Igarashi2011}
The RIXS spectra have been expressed as the one-spin correlation function
in the channel with changing polarization and as the two-spin correlation
function in the channel without changing polarization.
We have demonstrated that the contribution of the two-spin correlation 
function could be observed from the polarization analysis to possible RIXS 
spectra in Sr$_2$CuO$_3$.

\begin{acknowledgments}
We thank Professors M. Grioni and H. M. R\o nnow for valuable discussions.
This work was partially supported by a Grant-in-Aid for Scientific Research 
from the Ministry of Education, Culture, Sports, Science and Technology
of the Japanese Government.

\end{acknowledgments}

\appendix
\section{Absorption coefficient}
\begin{figure}
\includegraphics[width=8.0cm]{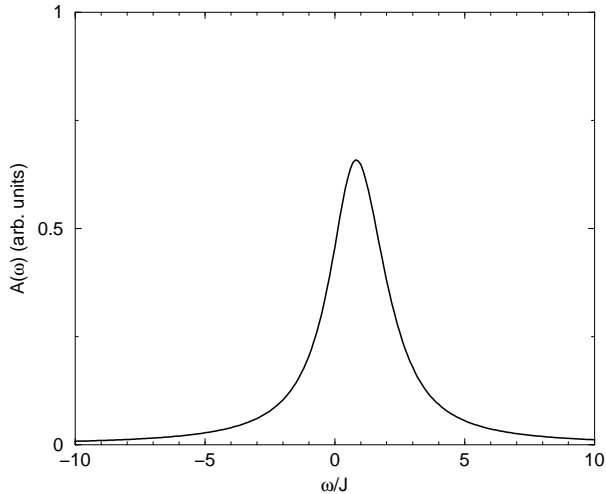}%
\caption{\label{fig.abs}
Absorption coefficient $A(\omega_i)$ as a function of photon energy $\omega_i$.
$\Gamma/J=1.2$. The origin of energy is set to correspond to
$\omega_i=\epsilon_{\textrm{core}}$. 
}
\end{figure}

The $L_{2,3}$-absorption coefficient $A_j(\omega_i)$ ($j=3/2$ or $j=1/2$)
close to $2p\to 3d_{x^2-y^2}$ transition may be given by the formula,
\begin{equation}
 A_j(\omega_i) \propto 
  \sum_{\sigma,\eta}|\langle \phi_{\eta}|\psi_0^{\sigma}\rangle|^2
  \frac{\Gamma/\pi}{(\omega_i+\epsilon_g-\epsilon_{\textrm{core}}
  -\epsilon'_{\eta})^2+\Gamma^2}.
\label{eq.abs}
\end{equation}
By substituting the eigenvalues and the eigenstates evaluated on finite-size
chain into Eq.~(\ref{eq.abs}), we obtain $A_{j}(\omega_i)$.
Figure \ref{fig.abs} shows the calculated $A_{j}(\omega_i)$
as a function of photon energy. 
The origin of photon energy is set to be $\omega_i=\epsilon_{\textrm{core}}$, 
and $\Gamma/J=1.2$.
The calculated curve is found very close to the Lorentzian shape.
The peak position is slightly shifted from $\omega_i=\epsilon_{\textrm{core}}$;
$\omega_i=\omega_i^0=\epsilon_{\textrm{core}}+0.8J$ for $\Gamma/J=1.2$.

\bibliographystyle{apsrev}
\bibliography{paper}

\end{document}